\documentclass{preprint}

\usepackage{moreverb,url}

\usepackage[colorlinks,bookmarksopen,bookmarksnumbered,citecolor=red,urlcolor=red]{hyperref}

\newcommand\BibTeX{{\rmfamily B\kern-.05em \textsc{i\kern-.025em b}\kern-.08em
T\kern-.1667em\lower.7ex\hbox{E}\kern-.125emX}}

\begin{document}

\runninghead{}
\runninghead{Organisciak and Ryan}

\title{Improving Text Relationship Modeling with Artificial Data}

\author{Peter Organisciak and Maggie Ryan} 

\affiliation{{\large\textbf{Peter Organisciak}} \\
            University of Denver, peter.organisciak@du.edu  \\
            \vskip 5pt
            {\large\textbf{Maggie Ryan}} \\
            University of Denver, maggie.ryan@du.edu}

\corrauth{Peter Organisciak,
University of Denver,
1999 E Evans Ave,
Denver, CO
80208 USA.}

\email{peter.organisciak@du.edu}

\begin{abstract}
Data augmentation uses artificially-created examples to support supervised machine learning, adding robustness to the resulting models and helping to account for limited availability of labelled data. We apply and evaluate a synthetic data approach to relationship classification in digital libraries, generating artificial books with relationships that are common in digital libraries but not easier inferred from existing metadata. We find that for classification on whole-part relationships between books, synthetic data improves a deep neural network classifier by 91\%. Further, we consider the ability of synthetic data to learn a useful new text relationship class from fully artificial training data.

\keywords{digital libraries, machine learning, data augmentation.}

\end{abstract}

\maketitle

\section{Introduction}
Identifying whole/part relationships between books in digital libraries can be a valuable tool for better understanding and cataloging the works found in bibliographic collections, irrespective of the form in which they were printed. However, this relationship is difficult to learn computationally because of limited ground truth availability. In this paper, we present an approach for data augmentation of whole/part training data through the use of artificially generated books. Artificial data is found to be a robust approach to training deep neural network classifiers on books with limited real ground truth, working to prevent over-fitting and improving classification by $91.0\%$.

Modern cataloging standards support encoding complex work-level relationships, opening the possibility for bibliographic collections that better represent the complex ways that works are changed, iterated, and collated in library books. Traditionally in bibliography, cataloging has controlled for work relationships only at the level of exact, so-called \textit{manifestation}-level duplicates, where the same work is represented identically in its physical form, and cataloging at higher levels of sameness - such as reprints or new editions - is a challenging manual process. Recent work has begun looking at computer-assisted ways of learning more about work relationships and properties \cite{bamman_estimating_2017,organisciak_characterizing_2019}, capitalizing on computational access to books afforded by large digital libraries.

\begin{figure}
\includegraphics[width=.45\textwidth]{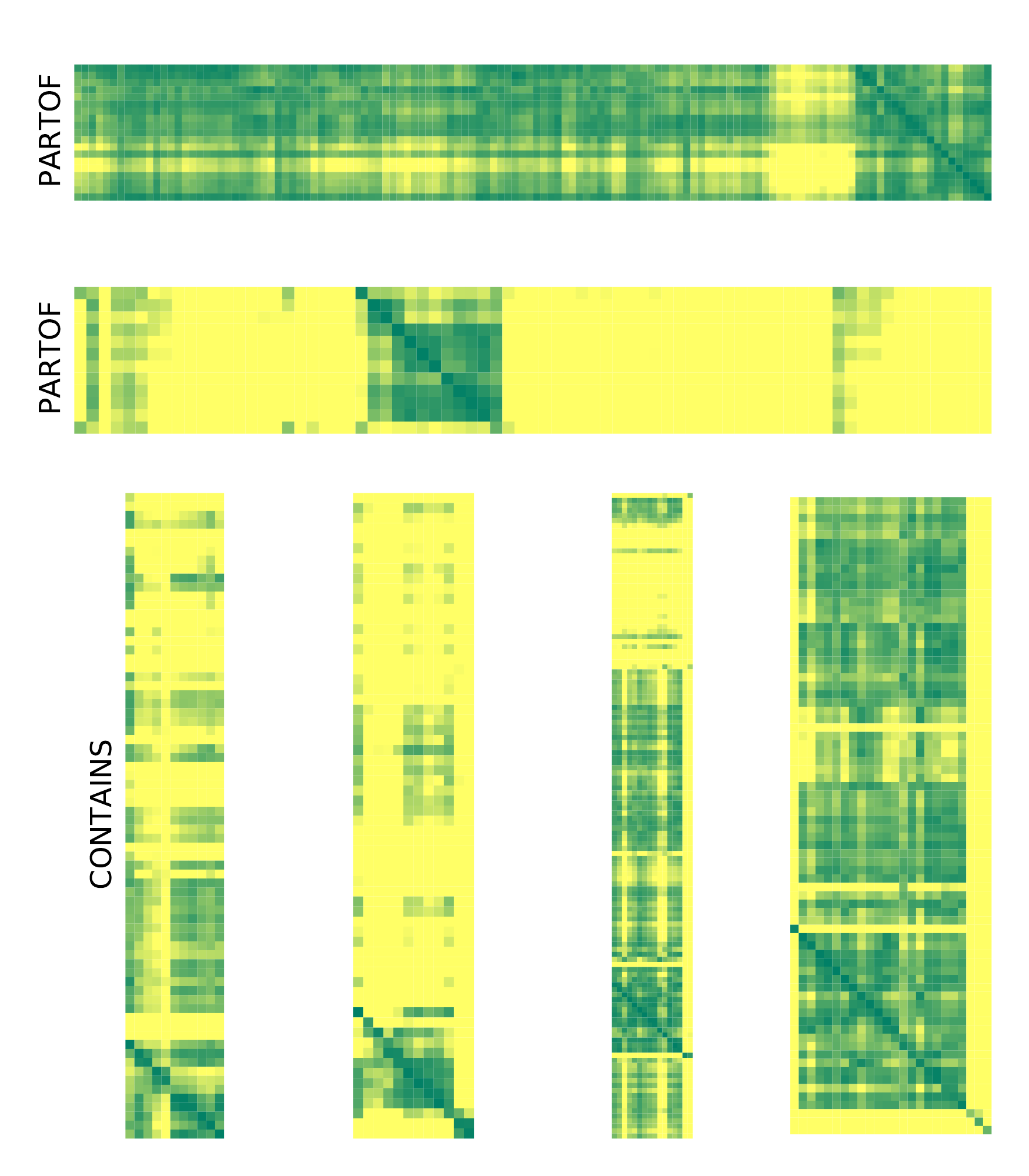}
\caption{What the classifier sees: examples of book-to-book similarity matrices on PARTOF and CONTAINS. Darker squares denote greater similarity between sub-units of the two books, and alignments of identical or extremely similar content are generally recognizable through a prominent diagonal pattern.}
\label{fig:simmat}
\end{figure}

Learning to tag work relationships from scanned digital library volumes relies on existing ground truth to learn from. Since granular relationships have not been historically cataloged, ground truth needs to be inferred from metadata using a mix of heuristics \cite{gatenby_glimir_2012} or additional authoritative sources, such as OCLC. One type of relationship that is difficult to infer is a whole-part relationship between works, where two books do not have the identical content, but overlap partially or one is subsumed by the other. Figure~\ref{fig:simmat} visualizes the pairwise similarity between books that hold this type of relationship, where the dark red diagonal of identical or near-identical content fully spans one axis (i.e. one book) but not the other. Examples of this relationship include:

\begin{itemize}
\item Anthologies and the individual works held within them, where a number of smaller works are contained within a larger one.
\item Anthologies and other anthologies, where there may be overlap between sub-works within them.
\item Abridgments and unabridged books, where the abridged version removes or shortens the full text.
\item Multi-volume sets vs a single-volume printing, where in one instance a full work is published in three books, where in another, it is published in just one.
\end{itemize}

A proper tagging of whole-part relationships in digital libraries presents promise for improving information access in our digital libraries, making it more clear which materials are within a multi-work book or for completing collections that are spread across multiple volumes. It also is immensely valuable for scholarship that tries to learn about language or culture by applying text mining to digital libraries, allowing text models to better avoid repeating texts and works, which may otherwise confound an accurate analysis of a collection \cite{schofield_quantifying_2017}.

We demonstrate that, even with a small number of training examples for whole-part relationships, it is possible to improve classification results through the use of artificially generated book data to support training. Successfully generating training examples requires a case where an expert can effectively describe the structure of the synthetic data, and may not apply for all instances. Yet, our strong results suggest that for cases where the characteristics of a class can be sufficiently described, synthetic data can be feasible way to communicate those properties to an text-based neural network classifier. We briefly consider that possibility in practice: after observing synthetic data greatly improve classification on low-support training labels, we train a classifier on a fully artificial -- but useful -- class of partially overlapping books. On manual review, we see that our artificial 'overlaps' class surfaces metadata and cataloguing errors from a test set of labeled examples. Leveraging synthetic data on such low- or no-support classes has the potential to help identify more scanning, printing, or metadata errors in digital libraries.

\subsection{Background}

This work grapples with a dilemma of computationally-aided cataloguing: while algorithmic approaches have the potential to assist in identifying historically under-catalogued bibliographic properties, that same lack of existing information complicates the ability to teach those algorithms. Specifically, we focus on \textit{work} relationships, a relatively recent affordance (\cite{american_library_association_rda_2013}) in cataloguing standards that makes it easier to control for when two books are different printings, editions, or parts of a larger whole. This paper looks at work relationship classification in the HathiTrust Digital Library, a digital library of more than 17 million scanned books from a worldwide consortium of library and memory institutions. Given that the majority of these materials are professionally catalogued, it also means that the collection is a very rich source for supervised learning. At the scale of the HathiTrust, even metadata that is spotty or inconsistent can usually still offer plenty of support for training and evaluating classifiers. However, this availability does not apply over all important work relationship labels, so we evaluate whether artificially generated books can fill the gap.

The most common contemporary way of thinking of different levels of sameness in collections is the four-part hierarchy of relationships for class one entities outlined by the Functional Requirements for Bibliographic Records (FRBR): \textit{work}, \textit{expression}, \textit{manifestation}, and \textit{item} \cite{ifla_study_group_on_the_functional_requirements_for_bibliographic_records_functional_1998}. A work is described as a distinct intellectual or artistic creation. An expression is the specific intellectual or artistic form that a work takes each time it is realized. Two different expressions of a work may be different editions or tellings of the story. The manifestation is the physical embodiment of an expression of a work. As an entity, manifestation represents all the physical objects that bear the same characteristics, in respect to both intellectual content and physical form. Lastly, an item is the single exemplar of a manifestation. This is the concrete entity; The physical item you can hold in your hand. These relationships allow for cursory identification of related works and fail to address the abstractions between expressions, manifestations, and items to support modern cataloging standards. Bibliographic materials are usually controlled at the manifestation-level, such as through ISBNs. However, the current Resource Description and Access (RDA) cataloguing standard implements the entities from FRBR, allowing for this form of granularity to be catalogued.

Improving bibliographic records opens up the possibility for information professionals to improve or correct metadata based on its relationship to other works, improve access and discovery within collections, develop a semantic relationship between collections and resources, and allows for the identification of the most representative, or cleanest, copy of a specific work within a collection. Additionally, the identification of stylistic similarities can provide valuable information for identifying contextually related but different works, assisting the process of discovery of materials and impact collection development by identifying gaps or overages of information within a collection.

In this paper, we have five working relationships to assist in identifying relationships between works: \texttt{SW}, same works, regardless of whether they are the same or a difference manifestation, \texttt{DV}, different volumes of the same overarching work; \texttt{CONTAINS}, where one book contains the work of the second book within it; \texttt{PARTOF}, where a book is fully contained within another book's work, and \texttt{DIFF}, referring to different work relationships. This is a streamlining of various possible same- or related-work relationship classes. A more detailed definition might differentiate between different expressions or manifestations of the same work, or include more detailed different work relationships, such as works by the same author.

Our focus is on the \texttt{PARTOF} and \texttt{CONTAINS} classes. These labels represent a 'whole-part' relationship between works, where two distinct works are related in that on subsumes the other. For example, an anthology of \textit{The Works of Charles Dickens} is considered a work in it's own right, but it may be comprised of other works, like \textit{Oliver Twist}. Identifying such relationships can better inventory all the works seen in a digital library, in the service of improved information access and retrieval. Scholars applying text analysis to study such collections may benefit from a better ability to avoid duplicated -- and potentially misleading \cite{schofield_quantifying_2017} -- text. Understanding the composite parts of anthologies and collections may even be valuable for finding works that inversely are not published alone.

Whole-part relationships are difficult to infer from metadata, since they usually are not directly catalogued. There are a few potential sources which may suggest the relationship, with limited coverage. First is volume enumeration information, which is held in the \texttt{995v} field of MARC metadata records. When used, this field describes the part, issue, or volume of a larger set that the given book is. For example, a printed or scanned book may be 'v.6' of a seven-part set. Sometimes multiple volumes are bound together, so a book that contains 'v.6-9' can be assumed to contain multiple volumes of a larger work.

Another potential bibliographic metadata source looks inward, enumerating the contents with a printed or scanned book. The MARC \texttt{500} field provides general notes, which may contain content information, and the \texttt{505} field provides formatted contents notes such as table of contents. Unfortunately, use of the \textit{5xx} fields are inconsistently applied so we cannot rely on this field alone to provide contents information for materials. In the HathiTrust, just over 2\% of all volumes have 505 information.

With the 995 field offering few examples of whole-part relationships and 505 fields difficult to parse and align, it is difficult to find sufficient training examples for training this type of relationship; in our data, real labels for whole-part data comprise only $0.63\%$ of training data. To enhance the reliability of our ground truth, we incorporate synthetic data, generating books that are artificially stitched together, applying an approach that has shown value in other domains.

\section{Related Work}

The past decade has seen an emergence of digital libraries at unprecedented scale, such at the Google Books project \cite{michel_quantitative_2011}, Internet Archive, and the HathiTrust \cite{york_building_2010}. This collections, comprised primarily of scanned books and other materials, are notable among text corpora for a number of reasons. One is the issues inherent to digitized texts, such as OCR errors \cite{smith_research_2018}. However, these corpora also span decades or centuries, at scales that have previously only been seen with contemporary text collections. Additionally, many texts originate in libraries, particularly academic libraries, and are accompanied by professionally catalogued metadata.

Emerging areas of study make use of the historical breadth and depth of large bibliographic collections to learn from digital library content at scale. The area of \textit{culturomics} seeks to infer aggregate-level trends about the emergence of history, culture, or language (\cite{michel_quantitative_2011}), such as work into changing meanings of words \cite{hamilton_diachronic_2016} or the regularization of irregular verbs \cite{lieberman_quantifying_2007}. In the digital humanities, scholars work to augment the traditional close reading exercises of literary appreciation with \textit{distant reading} \cite{moretti_distant_2013} and \textit{cultural analytics} \cite{manovich_cultural_2009}. Finally, in library and information science, large bibliographic collections allow us to effectively zoom out from a focus on individual materials, and learn more about those materials by observing their relationships in a larger system. Large digital library analysis can teach us more about bibliography, such as expanding our understanding of subject access, cataloguing practice, or the history of the book. For example, collecting duplicate copies of the same-work can be a useful resource for determining data of first publication \cite{bamman_estimating_2017}, aligned duplicate passages can be used to improve OCR \cite{dong_multi-input_2018}, and algorithmic modeling of books can test and uncover challenges to conceptual modeling of books \cite{organisciak_characterizing_2019}. Our work is positioned in this area, developing methods to learn about cataloguing practice and, in turn, supplement it.

This paper presents an approach grounded in artificially-created data. Artificial data is training data that has not been directly measured. It can be fully generated, \textit{synthetic} data \cite{nikolenko_synthetic_2019}, or it can be created through perturbation of existing data, sometimes called \textit{data augmentation} \cite{mikolajczyk_data_2018,wong_understanding_2016}. It is used for regularization in machine learning; that is, as a strategy for preventing over-fitting when training a model. In the case of artificial data, it is particularly valuable for overcoming supervised learning problems where it is difficult to collect enough labels.

Artificial data is particularly common in computer vision, where perturbing data into a realistic form is generally tractable. For example, a basic technique to make an image classifier more robust is to take each single labeled example, and use it to make new examples by flipping the image horizontally, cropping it, adding noise, blurring or sharpening, or adjusting the white balance \cite{perez_effectiveness_2017,mikolajczyk_data_2018,you_adversarial_2019}. More advanced techniques include changing hairstyles of people or artificially reposing them \cite{wong_understanding_2016}, or doing style transfer to adapt the existing examples to the styles \cite{mikolajczyk_data_2018}. Fully synthetic data is also common in computer vision \cite{nikolenko_synthetic_2019}, including for creating images from 3D models \cite[e.g.][]{peng_learning_2015} or even from imagery in games like \textit{Grand Theft Auto} \cite{richter_playing_2016}.

In text-based domains, data augmentation is usually used to perturb existing data, such as by swapping words with synonyms or word embedding nearest neighbors \cite{wei_eda_2019}, and randomly inserting or deleting words or adding noise \cite{wei_eda_2019,vaibhav_improving_2019}. Our paper does not deal with artificial adjustments to text itself, but rather to the structure of a book, although we expect perturbing the text in the books we use may add additional robustness to the approach. Our problem can be considered one of synthetic data, because while we re-purpose and remix existing book into new configurations, the configurations are synthetically created, and not derivative of the limited number of known training examples that we have.

Data augmentation is also used for balancing label representation in a dataset. In these cases, Generative Adversarial Networks (GANs) are increasingly used \cite[e.g.][]{zhu_emotion_2018,antoniou_data_2018,frid-adar_synthetic_2018}. A GAN is a two-part generative process, which simultaneously trains a generator to create data examples, such as a fake image of an object, and a discriminator, which aims to differentiate generated examples from real examples. Their competition encourages further growth - a generator has to get better to fool the discriminator, and the discriminator has to get better to avoid being fooled. This approach to synthetic data still needs sufficient data to train the GAN, but it provides an opportunity to inspect those training examples with a specificity that a classifier might not have. GANs are notoriously unstable, and are difficult to apply in text contexts. Our work follows in the tradition of synthetic data for label balancing. While it does not use GANs, it approaches the generation process as one akin to images: sidestepping complex considerations of syntax in text use and instead focuses on the structure of pages and books while remixing real examples of text.

\section{Data}
For this study, we focus on the HathiTrust Digital Library, a massive, 17-million work digital library. The HathiTrust is a non-profit consortium of institutions collecting their scanned bibliographic works for preservation and access \cite{york_building_2010}. It is appropriate for our goals because, due to the distributed way in which its collection was build, it offers a great deal of overlapping or duplicate works in various forms. Where a single library may not acquire tens or hundreds of different versions of a work, this type of bifurcation is more likely in the HathiTrust given that it's sourced from independently-built collections.

The HathiTrust, through the HathiTrust Research Center, also provides various tools, services, and datasets to aid scholarship over the collection. For this paper, the HathiTrust Extracted Features Dataset is used \cite{organisciak_access_2017}, which provides token count information for each page of each book in the collection.

\section{Methods}
To evaluate the efficacy of artificial data for whole/part relationships, we incorporate it into identically instantiated deep neural network relationship classifiers, in various mixes alongside known relationships. The three conditions evaluated are:

\begin{itemize}
\item Baseline (\texttt{nofake}): a classification model trained without any artificial data,
\item Mixed Train (\texttt{mixed}): a model trained with a mix of artificial data and ground truth, and
\item Artificial-Only (\texttt{allfake}): a model trained with only artificial data representing the \texttt{PARTOF} and \texttt{CONTAINS} classes.
\end{itemize}

In each condition, the classifiers are trained on the following classes:  The classes used for this paper are the two whole/part classes of interest to this paper - \texttt{PARTOF} and \texttt{CONTAINS} - as well as same-work, different volume of the work, and different work. The additional classes are used to help understand classification errors. 

The \texttt{nofake} condition presents a baseline condition, representing a situation without artificial data. After withholding texts for testing, we have a total of $11903$ training relationships for the two classes. For all other classes, we have 1.85 million labels, so the whole-part classes comprise $0.63\%$ of all training data.

The \texttt{mix} condition adds $272475$ artificially generated documents to the training set. This is an ideal parameterization, using artificial data to augment a class with disproportionately few training examples for supervised learning. With the synthetic data included, the whole-part examples comprise $12.74\%$ of all training labels.

Finally, the \texttt{allfake} condition serves to provide an impression of how successfully the artificial data represents the class. While it is not a realistic practical approach to actively remove the real training examples, even if they are underrepresented, it offers us a glimpse into how successfully the generated anthologies 'pretend' to be real ones. In this condition, we still evaluate on the same known relationships as in the other conditions, even if we are not training with them.

\subsection{Ground truth}
In each case, we evaluate the model's performance on the same split of held-out ground truth data ($n=12358$ for \texttt{PARTOF}, $n=12386$ for \texttt{CONTAINS}). 

Ground truth was extracted using metadata heuristics. Primarily, the volume enumeration field was used, which libraries contributing to the HathiTrust take from the MARC \texttt{995v} field or a number of other possible locations. This information lists which volume of a multi-volume work the given scan is. For pre-processing and information extraction, we first cleaned and normalized the field for 9 million English-language HathiTrust texts, standardizing atypical values, such as \textit{volume 1} or \textit{v1} instead of \textit{v.1}.

Subsequently, we used heuristics to identify when the chronology information listed a series for the given scan, extracted the individual parts of that series, and looked for corresponding volumes that only held those individual parts. For example, when a HathiTrust scan was listed as containing volumes 2 and 3, just volume 2 of the same book was searched for, as well as only volume 3. When there was an alignment, the relationship was saved as \texttt{CONTAINS} and the inverse relationship was tagged as \texttt{PARTOF}.

\subsection{Artificial data}

Our whole-part artificial data was created by breaking up or patching together existing books. Specifically, we create three types of artificial books:

\begin{itemize}
    \item Artificial anthologies, comprised of multiple short works by the same volume;
    \item Artificially combined volumes, created by concatenating scans which are different volumes of the same collection (e.g. combining v.1 and v.2 of Jane Austen's \textit{Emma}); and
    \item Smaller volumes artificially split from a longer book.
\end{itemize}

The short books used for anthologies were those that are shorter the than 40\% percentile of book lengths in the English-language HathiTrust. After de-duplication by author and title, this pool contained $757k$ books. During concatenation for anthologies and combined volumes, front matter and back matter were roughly separated by removing a randomly selected number of pages, up to 10, from the front and back of each book. The front and back of one of the sub-unit books was kept as the front and back matter of the subsequent larger book, with the center content of all the books included in the middle.

The classification setting here is relationship classification, which compares the relationship between two books. Thus, for each training example with synthetic data, the input is actually a comparison between a real book and the synthetic book that was constructed to contain or excerpt the real book.

\begin{table}
\caption{Training support for different classes.}\label{tab1}
  \begin{tabular}{rrl}
  \hline
  Label &  count \\
  \hline
    & \textit{ground truth} & \textit{synthetic}\\
    SW & 1029408 &\\
    DV & 376624 &\\
    PARTOF & 5908 &136204\\
    CONTAINS & 5995 &136271\\
    DIFF    &	460122 &\\
  \end{tabular}
\end{table}

\subsection{Deep neural network classification}
The robustness of artificial data in training relationship classes is evaluated \textit{in situ}, by including it in the context of a larger relationship classification pipeline. For training, we duplicated an architecture that worked well internally for other relationship classification. While this paper does not provide a deep treatment of of the architecture, the details are described below and in the supplemental code. Importantly, the same parameters and maintains in all experimental conditions.

For each left/right book pair, the classifier is given two inputs. The first input is a pairwise similarity matrix based on chunked sub-units of the book. Each book was collected into $5000$ word chunks. These were projected in a dense vector space, using a pre-trained 300-dimensional GloVe model \cite{pennington_glove_2014}, by converting each word to a vector and summing all the words for the chunk. With a single-summed word vector for each chunk, pairwise cosine similarity was performed between the chunks of both works, and the resulting similarity matrix was zero-padded or truncated to a size of $150$x$150$. For this study, we did not use texts that were truncated in the training data; that is, books which were more than 750k words.

While the first classifier input includes a sense of internal similarity between books, it does not capture topical differences. To do that, the second classifier input was derived from an overall GloVe vector for each book. The left book and right book vectors were used to calculate a centroid vector between the two, which was concatenated with the difference vector where the right was subtracted from the left.

The classifier used the similarity matrix input in a 2-dimensional convolutions neural network (CNN), extracting convolutions in a moving window of the similarity matrix, akin to how adjacent sequences of pixels are looked at for patterns in an image CNN. Two CNN layers are used, with down-sampling through maximum pooling, which drops information to avoid over-fitting and to reduce the parameter count. The parameters from the CNN are flattened, and dropout - a technique that randomly hides nodes on training to encourage more creative models - is applied \cite{srivastava_dropout_2014}. For the second input, a typical multi-layer model is used, also leveraging Dropout, and the weights are concatenated with the weights of the similarity matrix layer.

An example of how the classifier sees the book input for whole-part relationships is shown in Figure~\ref{fig:simmat}. Here, each book is split into 5000-word chunks, those word chunks are projected to a 300-dimensional vector in a word embedding space, and the figure depicts the cosine similarity between each chunk-to-chunk comparison across two books. Darker squares denote greater similarity, and in each, you can track the darkest diagonal to see pages of overlapping content. For \texttt{PARTOF} and \texttt{CONTAINS}, the diagonal spans the length of one axis but not the other.

\section{Results}

Tables~\ref{partofperformance} and \ref{containsperformance} shows the results of the respective conditions for including synthetic data. Overall, the synthetic data greatly improves the performance on the classes, with the macro-averaged $F_1$ Score improving from $0.411$ to $0.783$, a 37-point improvement (+91\%). Across all classes, the micro-averaged $F_1$ Score, which equally weighs each class, improved from $0.689$ to $0.815$.

The improvement to the classification $F_1$ Score comes through a massive 50-point gain to recall at a small 16-point cost to precision. In other words, without the synthetic data, the classifier rarely classified whole-part relationships except when it was very confidence -- missing many in the process. The synthetic data was not a perfect representation of what real data looks like, but it encouraged the classifier to better trust the general rules that were encoded in the heuristics driving the synthetic data generation.

When trained without \textit{any} real data, the synthetic-only data still outperformed the regular condition without data, though not to the level of the full, mixed-input. Still, it is notable that the fully synthetic data communicated something about the structural identity of class it was imitated. At the same time, given that the synthetic data outnumbered the real data $22:1$, the fact that the mixed class improved over the synthetic-only class -- both in precision and recall -- shows that it was still using the ground truth to aid in cases that were overlooked by the synthetic generation criteria.

\begin{table}
\caption{Performance on \textit{PARTOF} relationships by classifiers trained with varying mixes of artificial data.}\label{partofperformance}
  \begin{tabular}{@{}llll@{}}
                                        Class &  Precision & Recall & $F_1$ \\
                                        \hline
                                        NOFAKE (baseline)& \textbf{0.96}	&0.29	&0.44		 \\
                                        ALLFAKE &  0.75&	0.66&	0.70 \\
                                        MIX & 0.82&	\textbf{0.76}&	\textbf{0.79} \\
                                        \end{tabular}
\end{table}

\begin{table}
\caption{Performance on \textit{CONTAINS} relationships.}\label{containsperformance}
  \begin{tabular}{llll}
                                        Class &  Precision & Recall & $F_1$ \\
                                        \hline
                                        NOFAKE (baseline) & \textbf{0.97}&	0.24&	0.38 \\
                                        ALLFAKE & 0.78&	0.64&	0.71 \\
                                        MIX & 0.80&	\textbf{0.77}&	\textbf{0.78} \\
                                        \end{tabular}
\end{table}

\begin{figure*}
\includegraphics[width=\textwidth]{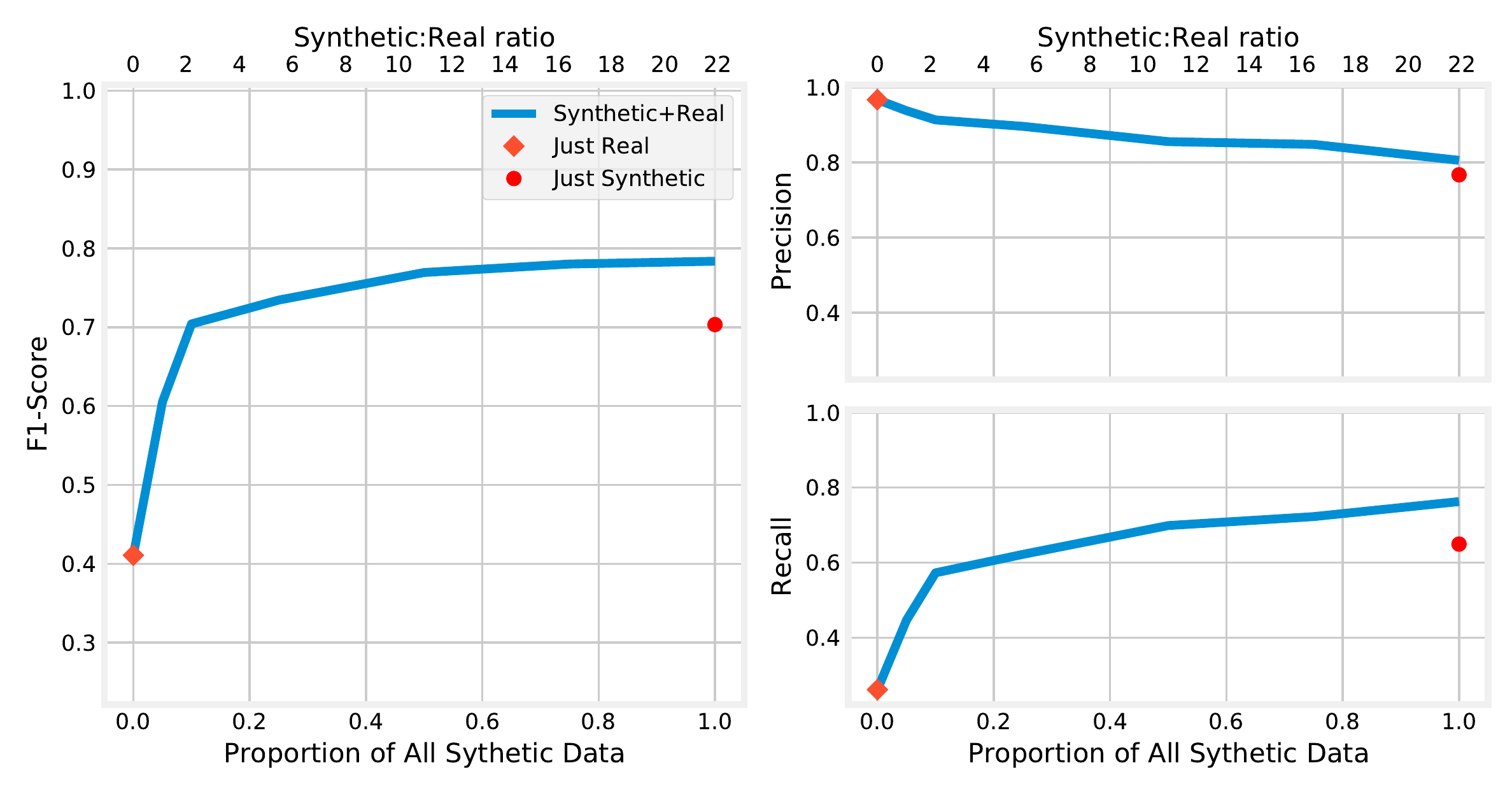}
\caption{Effect of varying amounts of synthetic data on classification of \texttt{PARTOF} and \texttt{CONTAINS} classes, where $n=272475$ with 100\% of synthetic data. Also shown are results when training \textit{only} on 100\% of synthetic data, with real labels removed from training. Labeled along the top is the associated \texttt{synthetic:real} ratio.}
\label{splits}
\end{figure*}

To consider whether too much synthetic data can be detrimental, we trained classifiers with different proportions of the synthetic data: 0\%, 5\%, 10\%, 25\%, 50\%, 75\%, and 100\%. Figure~\ref{splits} shows the effect of each on precision, recall, and $F_1$. We observe that $F_1$ keeps improving, with the majority of the improvements early. Underlying the $F_1$ Score improvement, there's again a gradual degradation in precision compensated by a larger improvement in recall.

\subsection{Overlapping books}
These results are promising for considering another class of relationship, for which very little ground truth is available: books that have a partial overlap. Is this section, we consider whether synthetic data can be used to surface a fully constructed relationship.

Consider the following two books: \textit{The writings of Oscar Wilde} (Wm. H. Wise \& Co., 1931), and \textit{The portable Oscar Wilde} (Penguin, 1981). Both volumes contain \textit{A Picture of Dorian Gray}, but they differ in the other content published within them.


The reason that there is little ground truth available for this type of overlapping relationship is that there are simply few use cases for it - it is valuable to know that \textit{Dorian Gray} is in an anthology, but not necessarily that two different anthologies contain it. However, this relationship is important because it is one that a classifier may encounter, and not understanding the \texttt{overlaps} relationship will cause a mis-classification when two such works are compared.

Overlapping work may be partially alignable from tables of contents, such as those sometimes listed in the MARC 505 field. Given its structural similarity to PARTOF and CONTAINS relationships, the results above suggest that a simulated data approach may also be tractable in this case. For a preliminary exploration, we developed synthetic overlapping data in the same manner as the whole-part data. We created fake anthologies, but while they shared material, this time neither fully subsumed the other.

\begin{table*}
\caption{Works most confidently classified as \texttt{OVERLAPS} by classifier trained on generated anthologies, shows as left (first item) and right (second item) pairs. Focus is only on public domain work, to allow manual review. HathiTrust scans are accessible through the persistent identifier http://hdl.handle.net/2027/\textbf{htid}.}\label{tab:overlapsbooks}

\begin{tabular}{p{.1cm}lllll}
 &  ground truth &           left & right &                                         title \\
1& CONTAINS &  uc1.31210024794651 &  uc1.32106007303982 &  Foreign direct investment in the United States... \\
  2 &  SW &  nyp.33433081888137 &      umn.31951000747566r & The Madras journal of literature and science. \\
3 &CONTAINS &     nnc1.cu04802497 &  uc1.c026291064 & List of the specimens of the British animals i... \\
4&CONTAINS &  mdp.39015017379614 & uiuo.ark:/13960/t3b000s6x &            High school mathematics, first course. \\
 5&CONTAINS &  msu.31293105610012 & uc1.b4687781 &                            Churches of Yorkshire. \\
  6 &PARTOF &  mdp.39015055357084 & mdp.39015037728535 &  Professional paper / Department of the Interio... \\
 7 &   SW &  nyp.33433090914031 & uiug.30112088599110 & Practical shop work; a concise explanation of ... \\
 8 &  DV &  uc1.31175010467416 &nnc1.0315301753 &  The works, literary, moral and philosophical o... \\
 9 & PARTOF &        uc1.b4687781 &                             msu.31293105610012 &                             Churches of Yorkshire. \\
10&  DIFF &  hvd.32044097007140 & hvd.32044097007397 &      Intermediate arithmetic, by Charles W. Morey.\\
& & &&  Elementary arithmetic / by Charles W. Morey.\\

\end{tabular}
\end{table*}

\begin{figure*}
\includegraphics[width=\textwidth]{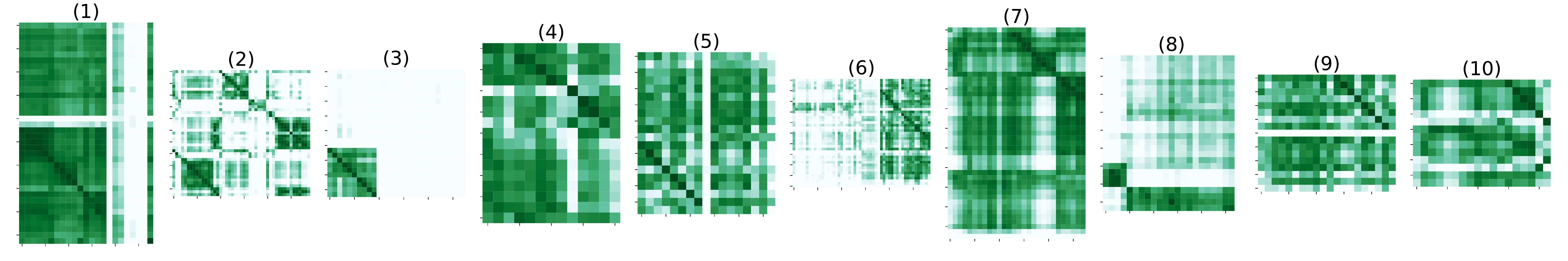}
\caption{Example relationships tagged as OVERLAPS, for the books depicted in Table~\ref{tab:overlapsbooks}, in the pairwise similarity matrix style of Figure \ref{fig:simmat}.}
\label{fig:overlapsmat}
\end{figure*} 

Table~\ref{tab:overlapsbooks} shows a selection of books that the resulting classifier considered to highly resemble the artificial \texttt{OVERLAPS} label. These are books from the evaluation performed earlier, meaning that they all already have a (different) ground truth label already. We manually reviewed the scans of the books lists, finding that most were indeed ground truth errors, resembling the intended OVERLAPS class. Associated similarity matrices for each listed example are shown in Figure~\ref{fig:overlapsmat}. For illustrative purposes, we consider three examples more in-depth below.


\begin{itemize}
\item \textit{The Madras journal of literature and science} (Left: nyp.33433081888137, Right:  umn.31951000747566r). For the ground truth, the metadata suggests that these books are identical: volume 16 of the Madras journal of literature and science (1850), with a similar number of pages. The reality is that both scans are incomplete, structurally resembling an OVERLAPS relationship. They start with \textit{no. 38} of volume 16; however, after the first page, the right book continues to page 2 while the left book jumps to page 174. After the truncated issue 38 ends in the left book, issue 37 is included. So the two scanned books intersect in front matter and the latter half of \textit{no. 38}, but diverge in that one has the first half also while the other has \textit{no. 37}.

\item  \textit{ List of the specimens of the British animals...} (Left: nnc1.cu04802497, Right: uc1.c026291064). For the ground truth, the metadata suggests that the left book is volumes 8 and 9 of \textit{List of the specimens of the British animals in the Collection of the British Museum}, and the right book is volume 9 -- a CONTAINS relationships. In reality, our classifier give this relationship a $94\%$ likelihood of being 'OVERLAPS'. The right book actually contains volume 9 of the listed work \textit{and} a second work, the \textit{Catalogue of the Genera and Subgenera of Birds Contained in the British Museum}.

\item \textit{High school mathematics, first course.} In the ground truth, the label is 'CONTAINS'. The left book is said to be volume 3 and 4, while the right book is said to be volume 3. The reality, however, is that right book has identical pages as the first half of the left book, but it has additional pages inserted throughout the book, interwoven without page numbers. 
\end{itemize}

This is a small exercise, looking at a small number of high confidence results. However, we also look at books that were already labeled -- books with unambiguous metadata that were deemed high-confidence enough to including in the evaluation corpus. Given that each was found to either be a misclassification or a printing error, this exercise suggests that there may promise in further pursuit of fully artificial classes.

\section{Discussion}

\subsection{Synthetic data is tractable for improving classification large text corpora}

The results reported here show a strong performance increase from the inclusion of synthetically-created books. The whole-part work relationships were the greatest source of mis-classifications in a larger classifier of work relationships, likely due to low relative representation in the collection. While their precision was high, there were many false negatives which, in a multi-label classifier, in turn lead to more false positives and lower precision for other classes. Through the inclusion of artificially-created books, the $F_1$ for the under-represented classes improved significantly.

When comparing different quantities of synthetic data, the results provide insight on how such strategies may be used in future digital library contexts. First, the synthetic data was beneficial even with a small set of training examples. Further, we found that increasing the quantity of examples -- which are trivial and computationally inexpensive to generate -- continued to improve the quality of results, albeit with gradually diminishing returns. 

We expect at some point that the synthetic data might start lowering the classifier accuracy as the real ground truth is overwhelmed, but we did not observe that point, even at a $22:1$ ratio of synthetic-to-real data. In this study, then, the exact quantity of synthetic data was found to \textit{not} be a highly sensitive parameter, and we expect that future use of the strategy would not require careful tuning.

Finally, we observed that while results were lower when real training labels were removed entirely, they still provided a notable improvement over the baseline. The strength of this condition was surprising, and seems to speak to the general strength of artificially concatenated books for our specific case.

\subsection{Generated data is appropriate for communicating class properties}

Why did the all-synthetic condition with real ground truth removed entirely outperform the baseline? One possibility is that the problem was particularly well-suited for describing a robust generator.

The characteristics of whole/part relationships are plain and uncomplicated enough for an expert-crafted imitation, where other types of labels may be more difficult to infer. A good classifier finds patterns that are too complex or too small for a person to observe. Yet, for all the internal complexity in deep neural network classification, this paper's results points to one way to 'talk' the language of the classifier when you do know how to describe a specific class. This is akin to how 3D or video game imagery can help train a computer vision classifier - a screenshot of a 3D-modelled streetlight may be missing some essence of a photograph, but will still convey many of the fundamentals of what a streetlamp looks like. In other words, if we have a label that can be described, writing a generator for synthetic data may allow us to nudge a high complexity classifier without sacrificing that complexity or requiring a re-engineering of the classifier. This promise seems supported by our brief look into teaching a new class label for overlapping data.

In information science, the possibilities for using fully-constructed book structures are intriguing. For example, one common type of printing and binding error is a book with runs of pages that have erroneously been included more than once, often in place of other, missing pages \cite{james_dawson_mistaikes_2016}. This type of error is imported into scanned book digital libraries, but as a structural issue, could one addressed through a synthetic data method akin to what we employed.

\subsection{Future directions}

Developing new synthetic classes in order to surface new types of relationship classes, or to train a classifier to identify books with printing or OCR errors is just one possible application of synthetic data in digital libraries.

The approach can be applied in various manners to augment classification approaches. In the domain of this paper, relationship tagging, we expect that comparing books with artificially-perturbed versions of themselves can aid in training same-work, same manifestation relationships - scans of books that have the same physical form, but where the text may vary due to OCR errors or marginalia. Replacing random words in a target text with typographical errors may help train downstream classifiers to be more robust to this type of error. Certainly, the real ground truth functions to do this. However, we have observed that the metadata underlying the ground truth is occasionally incorrect, such as with the examples that our OVERLAPS classifier surfaced. With data augmentation, we may grow the size of the training corpus while also limiting possible damage from miscatalogued books leading to ground truth errors.

We expect that this type of perturbation may further improve performance for the whole-part relationships observed in this study. The approach presented here does not perturb the text \textit{within} the pages being remixed for synthetic books. Thus, at classification, the portion of the similarity matrix between the real and synthetic book that overlaps is a self-similarity matrix, cleanly reflected around the diagonally. Real relationships, such as those seen in Figures~\ref{fig:simmat} and \ref{fig:overlapsmat}, have much more noise, due to smaller inconsistencies between scans. Despite the strong improvement seen in this paper, it is possible that the results could be improved further with such a strategy.

Other types of same-work relationship may be augmented with perturbed data similarly, to ensure a known relationship. Books that have the same expression printed in different manifestations may potentially be simulated by artificially staggering the page breaks of a real book, while books with different expressions of the same work might be possible to mimic be comparing a book to a version of itself with random words swapped for synonyms, an approach that has show strength for general text classification tasks \cite{wei_eda_2019}.

\section{Conclusion}

In this paper, we employ synthetic data as a means of label-balancing text training data for classification. In the context of scanned book digital libraries, we artificially remix real book through selective splitting or concatenation of real books, toward augmenting whole-part book relationship classes for which there is a disproportionately low amount of real data available. In essence, for a real book we generate a fake relative for that book, and use the relationship between the two in a classifier.

We find that the synthetic data is extremely effective for improving classification for our labels. Growing the size of the synthetic data improves performance, though the process is fairly permissive, and we find that carefully tuning for quantity is generally unnecessary. Additionally, relationships derived with artificial data train a surprisingly robust classifier even when the real ground truth is withheld. Encouraged by these results, we train and briefly review a fully constructed class, with encouraging results.

The results of this paper hold promise for information access in digital libraries in a number of ways. Directly, they will aid in building better methods for identifying relationships between books in digital libraries. More broadly, they hold promise for progressing bibliographic research. Synthetic data provides us a way to mix described bibliographic properties with known, labeled data in deep neural network classification, offering a way to improve supervised learning problems in contexts with limited or imbalanced training data.

\begin{funding}
This project was supported by IMLS \#LG-86-18-0061-18.

\end{funding}

\bibliographystyle{SageV}
\bibliography{citations}
\end{document}